\documentclass[latex,12pt]{article}
\usepackage{amssymb,amsmath}
\usepackage{graphicx} 
\usepackage{mathrsfs}
\usepackage{comment}
\usepackage{tikz}
\usepackage{fullpage}
\usepackage[sort&compress,numbers]{natbib}
\usepackage[colorlinks,breaklinks,bookmarks,pdfauthor=
 Martín A. Mosquera,pageanchor,citecolor=blue]{hyperref}

\begin{document}
\bibliographystyle{unsrtnat}
\newcommand{\mr}[1]{\mathrm{#1}}
\newcommand{\mc}[1]{\mathcal{#1}}
\newcommand{\ms}[1]{\mathscr{#1}}
\newcommand{\mb}[1]{\mathbf{#1}}
\newcommand{\ben}{\begin{equation}}
\newcommand{\een}{\end{equation}}
\newcommand{\XC}{_{\mr{xc}}}
\newcommand{\HXC}{_{\mr{Hxc}}}
\newcommand{\HXCa}{_{\mr{Hxc},\alpha}}
\newcommand{\XCa}{_{\mr{xc},\alpha}}
\newcommand{\Ha}{_{\mr{H}}}
\newcommand{\s}{_{\mr{s}}}
\newcommand{\uf}{\mr{f}}
\newcommand{\uG}{\mr{G}}
\newcommand{\us}{\mr{s}}
\newcommand{\uP}{\mr{P}}
\newcommand{\up}{\mr{p}}
\newcommand{\uXC}{\mr{xc}}
\newcommand{\uHXC}{\mr{Hxc}}
\newcommand{\ud}{\text{d}}
\newcommand{\ui}{\text{i}}
\newcommand{\intdr}{\int \text{d}^3\mb{r}~}
\newcommand{\dernr}[1]{\frac{\delta {#1} }{\delta n(\mb{r})}}
\newcommand{\derN}[1]{\frac{\partial {#1}}{\partial N}}
\newcommand{\dss}{\displaystyle}
\newcommand{\nv}[1]{[n_{v,#1}]}
\newcommand{\snv}[1]{n_{v,#1}}
\newcommand{\KS}{_\mr{KS}}
\newcommand{\cc}{_{\mr{c}}}
\newcommand{\ccH}{_{\mr{c,H}}}
\newcommand{\cs}{_{\mr{c,s}}}
\newcommand{\cXC}{_{\mr{c,xc}}}
\newcommand{\cHa}{_{\mr{c,H}}}
\newcommand{\zi}{z_{\mr{i}}}
\newcommand{\zf}{z_{\mr{f}}}

\newcommand{\dbar}{\mathchar'26\mkern-9mu \delta}

\newtheorem{thm}{Theorem}
\newtheorem{cor}{Corollary}

\title{{\scshape On the Action Formalism of Time-dependent Density-functional Theory}}
\author{Mart\'in A. Mosquera \\
Department of Chemistry\\ Purdue University\\ 
West Lafayette, IN 47907, USA\\ \texttt{ mmosquer@purdue.edu}}


\maketitle


\begin{abstract}    

The Runge-Gross [E. Runge, and E. K. U. Gross, Phys. Rev. Lett., {\bfseries 52}, 997 (1984)]
action functional of time-dependent density-functional theory leads 
to a well-known \emph{causality paradox}, i.e., a perturbation of the electronic density in the future
affects the response of the system in the present. This paradox is known to be caused by an inconsistent
application of the Dirac-Frenkel variational principle. In view of the recent solutions to this problem, 
the action functional employed by Runge and Gross in their formulation of time-dependent density functional theory 
is analyzed in the context of the Keldysh contour technique. The time-dependent electronic density, as well as the concept of causality, 
are extended to the contour.
We derive a \emph{variational equation} that obeys causality and relates the exchange-correlation potential
with its kernel, and the functional derivative of the exchange-correlation action functional with respect 
to the density. It is shown that the adiabatic local-density approximation is a consistent 
solution of this equation and that the time-dependent optimized potential method 
can also be derived from it.
The formalism presented here can be used to find new approximations methods
to the exchange-correlation potential and avoid the causality dilemma.  
\end{abstract}

\section{Introduction}


Time-dependent density-functional theory (TDDFT) \cite{GK90,MG04,U12} establishes the time-dependent (TD) electronic
density as the primary object of study to understand the dynamics of molecular systems.
TDDFT is widely used to calculate spectroscopic properties of molecules and solids, 
specially when TD perturbation theory is not applicable \cite{BWG05}. However, 
TDDFT can also be used to study electronic excitations in the linear regime, or predict the electronic ground-state energy and density \cite{LG06,HG11}.
The foundation of TDDFT is the theorem of Runge and Gross (RG) \cite{RG84} stating that there is a
one-to-one mapping, given an initial state, between electronic TD densities and 
TD external potentials. 
Later \citet{L99} showed that it is possible to reproduce the TD electronic density 
of the system of interacting electrons by a system of non-interacting electrons, 
which makes possible the use of the TD Kohn-Sham (KS) equations.
Challenges in TDDFT include the correct description of charge transfer excitation \cite{T03,HG12},
electronic transport through a molecule connected to metallic leads under a
bias \cite{K10,KCBC08,BCG05}, high-order harmonic generation \cite{KSK92}, double excitations \cite{MZCB04}, 
van der Waals interactions, among others \cite{EFB09}.

In TDDFT, the prediction of the evolution of the electronic density is 
reformulated in terms of the TD KS equations, 
which are easier to solve than the time-dependent Schr\"odinger equation (TDSE). 
Moreover, every observable of the system can be expressed as a functional of the density 
because the wave function is a density-functional as well. 
However, \citet{RG84} proved that the one-to-one mapping between TD densities and TD potentials is valid  
under the restriction that the TD external potential is Taylor-expandable in terms of the time variable.
The question as to how vast the set of TD potentials (or TD densities) of the RG theorem should be is still
an open question. The proof shown by \citet{RL11} and \citet{RGPL12} suggests that the Runge-Gross theorem
can be extended to a wider set of TD potentials including those that are non-analytic in time. 
Despite it is known that this map exists and there is a formal procedure to construct 
KS potentials, it is still a challenge to 
calculate the TD potential from a given TD density \cite{B08,NRLO12,RGPL12}.

In ground-state density-functional theory (DFT), the exchange-correlation (XC) potential 
is expressed as the functional derivative of the XC energy functional with respect to the time-independent electronic density. 
In TDDFT, an analogous variational relation between the TD XC potential and its action functional has been sought for the last three decades. 
\citet{P78} first suggested that the Dirac-Frenkel action functional and its variational principle 
should be used in TDDFT. Later \citet{RG84} showed
that the Dirac-Frenkel action functional extended to TDDFT leads to identify the TD XC potential as  
a functional derivative of the XC action functional with respect to the density. 
However, it was later found by \citet{GDP96} that this gives rise to a \emph{paradox} in which
a variation of the density in the future induces a perturbation of the potential in the past. According to this, 
the inverse first order response functional would not be causal. \citet{GDP96} conjectured that the paradox
could be solved by incorporating the causality principle explicitly into the action formalism.  

To resolve this causality paradox several works have been
published. \citet{R96} introduced an action based on the
work of \citet{JK79} in quantum field theory. However, this formalism does not use the density as basic variable
but a transition density that can be negative-valued; this
quantity is unsuitable as a basic quantity for TDDFT.

\citet{L98} proposed a functional in the Keldysh contour
with similar properties to that of a free energy. 
This functional depends on a pseudo-density in the Keldysh space that reduces 
to the density of the system when the potential in the Keldysh space corresponds to a physical potential. Due 
to the symmetry properties of the first-order response function 
of the pseudo-density, the causality is restored when the density is 
mapped to the real-time regime. However, the van Leeuwen formalism requires expansion of the action functional in terms of Feynman 
diagrams, while the functional of RG does not require such expansion. Furthermore,
the operator used in this formalism for the pseudo-density is not 
Hermitian in general and thus the pseudo-density does not integrate to the total number of electrons of the system; except
when the density is physical.

Recently, \citet{V08} solved the causality paradox in real-time by showing that the source of the problem in the RG formulation 
is a boundary condition. He showed that only the initial condition is necessary 
in the Runge-Gross functional to recover the causality restriction in general, and
derived an expression for the XC potential that is causal. 

In this paper I review Vignale's solution of the causality paradox in real-time from the perspective of unitary propagation and later use 
Vignale's theory to extend the RG action functional to the \emph{Keldysh contour}. The RG action functional 
in the Keldysh space, unlike the van Leeuwen functional, does not require diagrammatic expansion and uses
an electronic density that is a causal functional of the potential in the Keldysh space. By the RG
theorem applied to the Keldysh space and under the assumption that the density is a \emph{strictly causal} 
functional of the potential in the contour, I show that a \emph{variational 
equation} relating the XC potential with the XC action functional
arises. This equation shows an explicit dependence on the \emph{memory} of the system through the XC kernel.
I show that the adiabatic local density approximation (ALDA) is consistent with this equation,
how the TD optimized potential method (TDOPM) can be obtained, and also how 
the ground-state XC potential can be recovered. 

\section{Real-time Analysis}
The Dirac-Frenkel variational principle provides a method to derive the TDSE and its approximations by 
finding a stationary value of the action functional:
\ben
W[\psi;v]=\int_{t_0}^{t_1}\ud t~\langle \psi(t)|\ui \frac{\partial}{\partial t}-\hat{H}[v](t)|\psi(t)\rangle~.
\een
In this work we consider Hamiltonians of the form: 
\ben\hat{H}[v](t)=\hat{T}+\hat{W}+\int \ud^3\mb{r} ~v(\mb{r}t)\hat{n}(\mb{r})~,\een
where $\hat{T}$ and $\hat{W}$ are 
the kinetic energy and electron-electron repulsion energy operators, respectively, and $\hat{n}(\mb{r})$ is the 
density operator.  
The Dirac-Frenkel functional is defined over a \emph{Hilbert space of antisymmetric wavefunctions} representing 
bound systems of $N$ electrons. The TD Schr\"odinger equation (TDSE) is thus obtained by setting
\ben
\delta_{\psi} W[\psi;v] = 0~.
\een
This variational principle supposes that $\delta \psi(t_0)=\delta \psi(t_1)=0$. The solution of this equation, denoted
as $\psi[v](t)$, is the solution of the TDSE:
\ben
\ui \frac{\partial}{\partial t} |\psi[v](t)\rangle =\hat{H}[v](t)|\psi[v](t)\rangle~.
\een
$|\psi[v](t)\rangle$ is said to be a $v$-representable ket in real-time, which
expressed in terms of the unitary evolution operator is:
\ben
|\psi[v](t)\rangle=\hat{U}[v](t,t_0)|\psi(t_0)\rangle
~,\een
where
\begin{equation}\label{evol2}
\hat{U}[v](t,t_0)=\hat{\mc{T}}\exp\Big(-\ui\int_{t_0}^t \ud s~ \hat{H}[v](s)\Big)~~.
\end{equation} 
Here $\hat{\mc{T}}$ is the time-ordering operator in real-time. In this work we interpret the integral in the above 
equation to be taken over the interval $[t_0,t)$, i.e.
\ben
\int_{t_0}^t \ud s~\hat{H}[v](s)\ud s=\lim_{\epsilon\rightarrow 0}\int_{t_0}^{t-|\epsilon|}\ud s~ \hat{H}[v](s)~.
\een
Note that the ket $|\psi[v](t)\rangle$ is a causal functional of the potential: It is determined by the potential $v$ at times 
less than $t$. We refer to this dependency on the potential (Eq (\ref{evol2})) as the \emph{strict causality assumption}. 

All observables of $\psi$ are also causal functionals of $u$. 
For example, the density of the system,
\ben
n[v](\mb{r}t)=\langle \psi(t_0)|\hat{U}^{\dagger}[v](t,t_0)\hat{n}(\mb{r})\hat{U}[v](t,t_0)|\psi(t_0)\rangle~,
\een
is determined by $v$ in the interval $[t_0,t)$ \cite{U12}. By the RG theorem, given a fixed 
initial state, the potential $v$ at times in $[t_0,t)$ uniquely determines $n$ in the interval $[t_0,t)$, and vice versa.
If we denote as $u[n]$ the external potential as a functional of the TD density, 
then a first order variation in $u$ is given by a variation of $n$ over the interval $[t_0,t)$:
\begin{equation}\label{uofn}
\delta u[n](\mb{r}t)=\int_{t_0}^t \ud t'~\int \ud^3\mb{r}'~\chi^{-1}[n](\mb{r}t,\mb{r}'t')\delta
n(\mb{r}'t')~~,
\end{equation} 
where 
\ben\chi^{-1}[n](\mb{r}t,\mb{r}'t')=\frac{\delta u(\mb{r}t)}{\delta n(\mb{r}'t')}~.\een 
This indicates that $\delta u(\mb{r}t)/\delta n(\mb{r}'t')$ for $t\le t'$ is not defined because it does not
contribute to the integral of Eq. (\ref{uofn}).
However, for convenience we set:
\ben\chi^{-1}[n](\mb{r}t,\mb{r}'t')=0\quad t\le t'~.\een

\citet{V08}, however, employing the evolution equation of the current, showed
that $\delta u(\mb{r}t)/\delta n(\mb{r}'t')$ is related to $\delta(t-t')$ and its
first and second order time-derivatives when $t=t'$. 
This result is obtained under two assumptions different
from ours: First, $u(\mb{r}t)$ is determined by $n(\mb{r}'t')$ for $t'\le t$. 
And secondly, the functional derivative of the \emph{stress} tensor with 
respect to the density vanishes at equal times. Our assumption avoids this singularity in $\chi^{-1}$ 
and will be used to simplify our calculations in the Keldysh space (section \ref{kelsection}).

Now let us consider the Runge-Gross action functional:
\begin{equation}\label{rg_func}
A_v[n]=\int_{t_0}^{t_1}\ud t~\langle\psi[n](t)|\ui\frac{\partial}{\partial t}-\hat{H}[v](t)|\psi[n](t)\rangle~,
\end{equation}
where $|\psi[n](t)\rangle=|\psi[u[n]](t)\rangle$, $v$ is some TD external potential, and $t_1>t_0$. Note that
the ket $|\psi[n](t)\rangle$ is causal, i.e., it is determined by $n$ in the interval $[t_0,t)$.
\citet{RG84}, based on the Dirac-Frenkel variational principle, imposed $\delta \psi(t_0)=\delta \psi(t_1)=0$ and 
\ben\frac{\delta A_v}{\delta n(\mb{r}t)}=0~,\een 
which leads to the following alternative form of the variational principle:
\ben\label{intact}
\frac{\delta B[n]}{\delta n(\mb{r}t)}-v(\mb{r}t)=0~,
\een
where $B[n]$ is the internal action:
\ben
B[n]=\int_{t_0}^{t_1}\ud t~\langle\psi[n](t)|\ui\frac{\partial}{\partial t}-\hat{T}-\hat{W}|\psi[n](t)\rangle~.
\een
If Eq. (\ref{intact}) were valid then we could assert that \cite{RvL12}:
\ben
u[n](\mb{r}t)=\frac{\delta B[n]}{\delta n(\mb{r}t)}
\een
Unfortunately, when the above function is further differentiate with respect to $n$,
\ben
\chi(\mb{r}t,\mb{r}'t')=\frac{\delta^2B[n]}{\delta n(\mb{r}'t')\delta n(\mb{r}t)}~,
\een
one finds an inconsistency because the above equation implies that $\chi(\mb{r}t,\mb{r}'t')\neq 0$ for $t<t'$. 
This is known as the \emph{causality paradox} \cite{GDP96}. 
The solution to the paradox was found by \citet{V08}, who pointed out that, according to the definition of $v$-representable wave-function, 
we can only set $\delta \psi(t_0)=0$ because a perturbation $\delta n(\mb{r}t)$, in general, will induce a response $\delta \psi(t_1)\neq 0$. 

The solution of \citet{V08} can be viewed as a direct implementation of the causality principle into 
the RG functional. For example, the internal action $B[n]$, using the TDSE, can be written as \cite{U12}:
\ben
B[n]=\int_{t_0}^{t_1}\ud t~\int \ud^3\mb{r}~ u[n](\mb{r}t)n(\mb{r}t)~.
\een
The density-functional $u[n]$ is causal by the RG theorem. If we differentiate the above functional 
with respect to the density and insert the result into the functional derivative of the RG action functional
we obtain
\begin{equation}\label{var_principle}
\frac{\delta A_v}{\delta n(\mb{r}t)}=u[n](\mb{r}t)-v(\mb{r}t)+\int_{t}^{t_1}\ud t'~\int \ud^3\mb{r}'~\chi^{-1}[n](\mb{r}t,\mb{r}'t')n(\mb{r}'t')~.
\end{equation}
Now let $n_v$ be the TD density corresponding to $v$, then:
\ben
\frac{\delta A_v}{\delta n(\mb{r}t)}\Bigg|_{n=n_v}=\int_{t}^{t_1}\ud t'~\int \ud^3\mb{r}'~\chi^{-1}[n_v](\mb{r}t,\mb{r}'t')n_v(\mb{r}'t')~.
\een
This last equation is an alternative form of the Vignale variational formulation that shows 
that $n_v$ is not a stationary value of $A_v[n]$. This is a consequence 
of constraining the wave-functions of the RG functional to be density-functionals of the form $\psi[u[n]]$.
\citet{RvL12} showed that not every TD wave-function can be associated with a TD external potential (or a TD density). Hence 
the domain of the RG functional is just a subset of the domain of the Dirac-Frenkel functional, explaining
why the RG and the Dirac-Frenkel functionals lead to different results. 

\section{Keldysh-space Analysis}\label{kelsection}

Let us add a super index $+$ or $-$ to the time variable $t$. The Keldysh contour, $\mc{C}$, is expressed
as $\mc{C}=\mc{C}^+\cup \mc{C}^-$, where $\mc{C}^+=[t_0^+,t_1^+]$ and $\mc{C}^-=[t_0^-,t_1^-]$. 
We denote $z$ as a variable in the contour $\mc{C}$, and let $\zi=t_0^+$ and $\zf=t_1^-$. The arrow 
of time in $\mc{C}$ points from $t_0^+$ to $t_1^+$ and from $t_1^-$ to $t_0^-$. Thus, any $z\in\mc{C}^-$ is
said to be later than any $z'\in \mc{C}^+$. If $z,z'\in \mc{C}^-$ we say that $z$ is later 
than $z'$ if $t(z)<t(z')$, where $t(z)$ is the real value of $z$. 
A ket in $\mc{C}$ is denoted as $|\psi\cc[u\cc](z)\rangle$, where
$u\cc(\mb{r}z)$ is some potential in $\mc{C}$. A physical potential in $\mc{C}$ is denoted as $\bar{u}\cc$ 
and it satisfies $\bar{u}\cc(\mb{r}t^+)=\bar{u}\cc(\mb{r}t^-)$. 
Thus a potential in real-time is 
mapped to the Keldysh space when setting $\bar{u}\cc(\mb{r}t^{\pm})=u(\mb{r}t)$ ($t^{\pm}$ we denotes
evaluation at $\mc{C}^+$ or $\mc{C}^-$). 

We now extend the unitary propagator $\hat{U}$ to the Keldysh space as follows:
\begin{equation}\label{pevol}
\hat{U}\cc[u\cc](z,\zi)=\hat{\mc{T}}_{\mc{C}} \exp\Big[-\ui\int_{\zi}^{z}\ud z'~\hat{H}\cc[u\cc](z')\Big]~,
\end{equation}
where $\hat{\mc{T}}_{\mc{C}}$ is the path-ordering operator in $\mc{C}$ (for example, 
$\hat{\mc{T}}_{\mc{C}}[\hat{B}\cc(z')\hat{A}\cc(z)]=\hat{A}\cc(z)\hat{B}\cc(z')$ if 
$z$ is later than $z'$). The Hamiltonian in the Keldysh space now reads
$\hat{H}\cc[u\cc](z)=\hat{T}+\hat{W}+\int \ud^3\mb{r}~ u\cc(\mb{r}z)\hat{n}(\mb{r})$. 
The integration over the pseudo-time is defined as:
\begin{equation}
\int_{\zi}^{z}\ud z'~f\cc(z')=\begin{cases}
\int_{t_0}^t \ud t' ~f\cc(t'^+),~~~ z=t^+\\
\int_{t_0}^{t_1}\ud t' ~f\cc(t'^+)+\int_{t_1}^t \ud t'~f\cc(t'^-),~~~ z=t^-.
\end{cases}
\end{equation} 
A $v$-representable ket in $\mc{C}$ is thus expressed as $|\psi\cc[u\cc](z)\rangle=\hat{U}\cc[u\cc](z,\zi)|\psi\cc(\zi)\rangle$,
where $|\psi\cc(\zi)\rangle=|\psi(t_0)\rangle$ is the initial state of the system. Note that $\psi\cc(z)$ does not 
depend on the potential $u\cc$ at later times than $z$. As in the real-time case, we assume that the end point 
of the integral in Eq. (\ref{pevol}) is not included; this can be considered as an extension of the strict
causality assumption to the Keldysh contour.
We define the density in $\mc{C}$ as \cite{D12}:
\begin{equation}
n\cc[u\cc](\mb{r}z)=\langle\hat{U}\cc^{\dagger}[u\cc](z,\zi)\hat{n}(\mb{r})\hat{U}[u\cc]\cc(z,\zi)\rangle~~,
\end{equation}
where $\langle \cdot \rangle=\langle \psi\cc(\zi)|\cdot|\psi\cc(\zi)\rangle$. To prove that there is a one-to-one mapping between $n\cc$ and $u\cc$, it is sufficient to notice
that $\psi\cc$ satisfies the Schr\"odinger equation in $\mc{C}^+$. Therefore, if the potential
can be expressed as a power series around $\zi$, then the RG theorem and its extension \cite{RL11}
including non-analytic potentials apply in this case.

Let us examine the action functional proposed by \citet{L98}, which reads
\ben
A_{\mr{vL}}[u\cc]=\ui \ln \langle \hat{U}\cc[u\cc](\zf,\zi)\rangle~.
\een
The functional derivative of this functional with respect to the
potential $u\cc$ yields the pseudo-density \cite{L98}:
\begin{equation}
n_{\mr{vL}}(\mb{r}z)=\frac{\langle \hat{U}\cc(\zf,z)\hat{n}(\mb{r})\hat{U}\cc(z,\zi)\rangle}{
\langle \hat{U}\cc(\zf,\zi)\rangle}~~.
\end{equation}
However, the above density is an average of the operator:
\ben\hat{n}_{\mr{vL,H}}(\mb{r})=
\hat{U}\cc(\zf,z)\hat{n}(\mb{r})\hat{U}\cc(z,\zi)~,\een 
which is not a Hermitian operator. Therefore
\ben
\int \ud^3\mb{r}~\hat{n}_{\mr{vL,H}}(\mb{r}z)=\hat{N}\hat{U}\cc(\zf,\zi)~, 
\een
where $\hat{N}$ is the particle-number operator.
This implies that $n_{\mr{vL}}$ does not integrate to $N$; except when the potential 
$u\cc$ is physical \cite{L98}. The density $n\cc$, on the other hand,
integrates to $N$ and is always positive.

It can be shown that the response function of the density 
in $\mc{C}$ is given by:
\begin{equation}\label{kres}
\chi\cc[u\cc](\mb{r}z,\mb{r}'z')=\frac{\delta n\cc(\mb{r}z)}{\delta u\cc(\mb{r}'z')}=-\ui\langle 
[\hat{n}\ccH[u\cc](\mb{r}z),\hat{n}\ccH[u\cc](\mb{r}'z')]\rangle~,
\end{equation}
where the Heisenberg representation of the density operator $\hat{n}(\mb{r})$ is 
\ben
\hat{n}\ccH[u\cc](\mb{r}z)=\hat{U}\cc^{\dagger}[u\cc](z,\zi)\hat{n}(\mb{r})\hat{U}\cc[u\cc](z,\zi)~.
\een
Eq. (\ref{kres}) is valid if $z$ is later than $z'$, and we set
$\chi\cc(\mb{r}z,\mb{r}'z')=0$ if $z'$ is later than or equal to $z$.

The inverse first order response function $
\chi^{-1}\cc[n\cc](\mb{r}z,\mb{r}'z')=\delta u\cc(\mb{r}z)/\delta n\cc(\mb{r}'z')
$, according to the RG theorem extended to the $\mc{C}$, must also satisfy causality in the contour, e.g.,
$\chi\cc^{-1}(\mb{r}z,\mb{r}z')=0$ if $z=z'$ or $z'$ is later than $z$.
When a physical potential is used, the Heinsenberg operators recover their usual form in real-time.
Therefore, we obtain a physical density $\bar{n}\cc(\mb{r}t^{\pm})=n(\mb{r}t)$. From Eq. (\ref{kres})
we can show that the first order response function satisfies the antisymmetry relationship:
\begin{equation}\label{xx1}
\chi\cc(\mb{r}t^+,\mb{r}'t'^+)\Big|_{u\cc=\bar{u}\cc}=-\chi\cc(\mb{r}'t'^-,\mb{r}t^-)\Big|_{u\cc=\bar{u}\cc}~,
\end{equation}
where $u\cc=\bar{u}\cc$ denotes evaluation at the physical regime. 
Note that $\chi\cc$ also satisfies $\chi\cc(\mb{r}t^+,\mb{r}'t'^+)=\chi\cc(\mb{r}t^-,\mb{r}'t'^+)$ 
and $\chi\cc(\mb{r}'t'^-,\mb{r}t^+)=\chi\cc(\mb{r}'t'^-,\mb{r}t^-)$ if $t>t'$ and
$u\cc=\bar{u}\cc$. 

The response of the density in the Keldysh space is \cite{L98}:
\begin{equation}
\delta n\cc[u\cc](\mb{r}z)=\int_{\zi}^{\zf}\ud z'\int \ud^3\mb{r}' 
~\chi\cc[u\cc](\mb{r}z,\mb{r}'z')\delta u\cc(\mb{r}'z')~.
\end{equation}
To obtain the response in real-time, the variation of a physical potential must satisfy 
$\delta \bar{u}\cc(\mb{r}t^+)=\delta \bar{u}\cc(\mb{r}t^-)=\delta u(\mb{r}t)$. Using the 
aforementioned properties of $\chi\cc$ to calculate the above integral, 
the response of the density turns out to be independent of the time location in the contour, i.e. 
$\delta n\cc(\mb{r}t^+)=\delta n\cc(\mb{r}t^-)=\delta n(\mb{r}t)$. Hence, it is determined
by: 
\begin{equation}
\delta n\cc[\bar{u}\cc](\mb{r}t)=\int_{t_0}^{t^{\pm}}\ud t'\int \ud ^3\mb{r}'
~\chi\cc[\bar{u}\cc](\mb{r}t^{\pm},\mb{r}'t'^+)\delta \bar{u}\cc(\mb{r}'t'^{+})~.
\end{equation}
This result allows us to identify the response in real-time $\chi(\mb{r}t,\mb{r}'t')$ as
$\chi\cc(\mb{r}t^{\pm},\mb{r}'t'^+)|_{u\cc=\bar{u}\cc}$ or $-\chi\cc(\mb{r}'t'^-,\mb{r}t^-)|_{u\cc=\bar{u}\cc}$,
which are causal. 
Exchanging variables in the integral of $\chi\cc\chi\cc^{-1}$ 
reveals that $\chi\cc^{-1}$ satisfies the same relationships of $\chi\cc$ regarding exchange 
of variables at physical densities. 

Let us extend the functional $A_v$ to the Keldysh space:
\begin{equation}\label{acden}
\mc{A}_{\bar{v}\cc}[n\cc]=\mc{B}[n\cc]-\int_{\zi}^{\zf} \ud z \int \,\ud^3\mb{r}\,n\cc(\mb{r}z)\bar{v}\cc(\mb{r}z)~,
\end{equation}
where 
\ben\mc{B}[n\cc]=\int_{\zi}^{\zf} \ud z\,\langle \psi\cc[n\cc](z)|\ui\frac{\partial}{\partial z}-\hat{T}-
\hat{W}|\psi\cc[n\cc](z)\rangle~,\een 
$\bar{v}\cc$ is some external physical potential, and $\partial f(z)/\partial z=\partial f(t^\sigma)/\partial t$, where $\sigma=+,-$. 
Vignale equation in this case reads:
\begin{equation}\label{bs}
\begin{split}
\frac{\delta \mc{B}}{\delta n\cc(\mb{r}z)}\Bigg|_{n\cc=\bar{n}_{\mr{c},\bar{v}_{\mr{c}}}}&-\bar{v}\cc(\mb{r}z)=\ui\langle 
\psi\cc(\zf)|\frac{\delta\psi\cc(\zf)}{\delta n\cc(\mb{r}z)}\rangle \Bigg|_{n\cc=\bar{n}_{\mr{c},\bar{v}_{\mr{c}}}}\\
&=\int_{z}^{z_{\mr{f}}} \ud z'\int \ud^3\mb{r}'~\bar{n}_{\mr{c},\bar{v}_{\mr{c}}}(\mb{r}'z')
\chi\cc^{-1}[\bar{n}_{\mr{c},\bar{v}_{\mr{c}}}](\mb{r}'z',\mb{r}z)~.
\end{split}
\end{equation}
The left hand side of the above equation corresponds to $\delta \mc{A}_{\bar{v}\cc}/\delta n\cc(\mb{r}z)$
evaluated at the density that yields $\bar{v}\cc$, $\bar{n}_{\mr{c},\bar{v}_{\mr{c}}}$. Additionally, the above 
equation also gives the functional derivative $\delta \mc{B}/\delta n\cc(\mb{r}z)$ for 
an arbitrary density $n\cc$; in this case, we replace $\bar{v}\cc$ by $u\cc[n\cc](\mb{r}z)$,
$\bar{n}_{\mr{c},\bar{v}_{\mr{c}}}$ by $n\cc$, and the inverse response function 
has to be evaluated at $n\cc$.

Let us introduce the KS action functional:
\begin{equation}
\mc{A}_{\mr{s},\bar{v}\cs}[n\cc]=\mc{B}_{\mr{s}}[n\cc]-
\int_{\zi}^{\zf}\ud z~\int \ud^3\mb{r}~n\cc(\mb{r}z)\bar{v}\cs(\mb{r}z)~~,
\end{equation}
where $\bar{v}\cs(\mb{r}z)$ is some effective external potential and
\ben\mc{B}_{\mr{s}}[n\cc]=\int_{\zi}^{\zf}\ud z\,\langle\Phi\cs[n\cc](z)|\ui\frac{\partial}{\partial z} -\hat{T}|
\Phi\cs[n\cc](z)\rangle~.\een
The  KS wave function is a Slater determinant of TD KS orbitals $\{\phi_{\mr{c},i}(\mb{r}z)\}$ that satisfy:
\ben
\ui\frac{\partial\phi_{\mr{c},i}}{\partial z}=\Big(-\frac{1}{2}\nabla^2_{\mb{r}}+u\cs[n\cc](\mb{r}z)\Big)
\phi_{\mr{c},i}(\mb{r}z)~,\een
where $u\cs[n\cc]$ is the KS potential that represents $n\cc(\mb{r}z)$.
Thus, if we differentiate $\mc{B}_{\mr{s}}$ with respect to the TD density we obtain:
\begin{equation}\label{bKohn-Sham}
\frac{\delta \mc{B}_{\mr{s}}}{\delta n\cc(\mb{r}z)}=u\cs[n\cc](\mb{r}z)+
\int_{z}^{\zf}\ud z'\int \ud^3\mb{r}'~n\cc(\mb{r}'z')\chi\cs^{-1}[n\cc](\mb{r}'z',\mb{r}z)~,
\end{equation}
where $\chi\cs^{-1}(\mb{r}z,\mb{r}'z')=\delta u\cs(\mb{r}'z')/\delta n\cc(\mb{r}z)$.

Recall the Hartree functional:
\begin{equation}
\mc{A}\Ha[n\cc]=\frac{1}{2}\int_{\zi}^{\zf} \ud z\int \ud^3\mb{r}\int \ud^3\mb{r}'\,\frac{n\cc(\mb{r}'z)n\cc(\mb{r}z)}{|\mb{r}-\mb{r}'|}~.
\end{equation}
Let us introduce the XC action functional:
\begin{equation}\label{expansion}
\mc{A}\XC[n\cc]=\mc{B}_{\mr{s}}[n\cc]-\mc{B}[n\cc]-\mc{A}\Ha[n\cc]~.
\end{equation}
Using Eqs. (\ref{bs}) and (\ref{bKohn-Sham}) it is found that the functional derivative of the XC action
functional can be expressed as:
\begin{equation}\label{vxcK}
\begin{split}
u\cXC(\mb{r}z)+\int_{z}^{\zf}& \ud z'\int \ud^3\mb{r}'n\cc(\mb{r}'z')[\chi\cs^{-1}(\mb{r}'z',\mb{r}z)
\\&-\chi\cc^{-1}(\mb{r}'z',\mb{r}z)]=
\frac{\delta \mc{A}\XC}{\delta n\cc(\mb{r}z)}~.
\end{split}
\end{equation}
Here $u\cXC(\mb{r}z)=u\cs(\mb{r}z)-u\cc(\mb{r}z)-u\cHa(\mb{r}z)$, where the Hartree potential is
$u\cHa[n\cc](\mb{r}z)=\int \ud^3\mb{r}'\,n\cc(\mb{r'z})/|\mb{r}-\mb{r}'|$. 
Now introduce the XC kernel $f\cXC(\mb{r}z,\mb{r}'z')=\delta u\cXC(\mb{r}z)/\delta n\cc(\mb{r}'z')$, which satisfies:
\begin{equation}
\chi\cs^{-1}(\mb{r}z,\mb{r}'z')=\chi\cc^{-1}(\mb{r}z,\mb{r}'z')
+\frac{\delta\cc(z-z')}{|\mb{r}-\mb{r}'|}+f\cXC(\mb{r}z,\mb{r}'z')~.
\end{equation}
The delta function in $\mc{C}$ space is defined such that
$\int_{\zi}^{\zf}\ud z' f\cc(z')\delta\cc(z-z')=f\cc(z)$. The KS response function and 
the XC kernel satisfy the same properties of $\chi\cc$ regarding exchange 
of variables.

In order to simplify Eq. (\ref{vxcK}), suppose that the density is physical, $n\cc=\bar{n}\cc$. 
This imposes that the XC 
potential is the same in both $\mc{C}^+$ and $\mc{C}^-$ spaces. For example, if 
$z=t^+$ then the integral in time can be split up into two integrals: 
The first one runs from $t^+$ to $t^-$, and the second one from $t^-$ to $t_0^-$. 
There is no contribution from the first integral due to the symmetry properties of 
$\chi\cc^{-1}$ and $\chi\cs^{-1}$ at physical densities. For the second integral we can 
use the antisymmetry relation to obtain in real-time that:
\begin{equation}\label{eq:vxc}
u\XC(\mb{r}t)+\int_{t_0}^{t} \ud t'\int \ud^3\mb{r}'~f\XC(\mb{r}t,\mb{r}'t')
n(\mb{r}'t') =
\frac{\dbar \mc{A}\XC}{\dbar n(\mb{r}t)}~,
\end{equation} 
where $u\XC(\mb{r}t)=\bar{u}\cXC(\mb{r}t^{\pm})$ and
\begin{equation}
\frac{\dbar \mc{A}\XC}{\dbar n(\mb{r}t)}=
\frac{\delta \mc{A}\XC}{\delta n\cc(\mb{r}t^{\pm})}\Bigg|_{n\cc=\bar{n}\cc}~~.
\end{equation}
Setting $z=t^-$ in Eq. (\ref{vxcK}) also leads to Eq. (\ref{eq:vxc}) when $n\cc=\bar{n}\cc$; 
for this reason we expressed the final result in real-time.
Because $f\cXC$ in the $\mc{C}$ space 
also has the same properties as $\chi\cc^{-1}$ we identify the XC kernel in real-time,
$f\XC(\mb{r}t,\mb{r}'t')$, as $f\cXC(\mb{r}t^{\pm},\mb{r}'t'^+)|_{n\cc=\bar{n}\cc}$,
or $-f\cXC(\mb{r}'t'^-,\mb{r}t^-)|_{n\cc=\bar{n}\cc}$. Thus, the XC kernel is causal in 
real-time. 

Given that we assumed that the response functions $\chi\cc$ and $\chi\cs$ are 
strictly causal in $\mc{C}$, the integral in Eq. (\ref{vxcK}) is taken 
over the interval $(z,\zf]$. This implies that the Hartree kernel 
$\delta\cc(z-z')/|\mb{r}-\mb{r}'|$ lies outside the integration limits and thus it has
no contribution to Eq. (\ref{eq:vxc}). Based on this,
the integral in Eq. (\ref{eq:vxc}) is carried out strictly over the 
past of $t$, i.e., $[t_0,t)$. Hence, our causality assumption avoids
singularities at equal-times and simplifies the transition to real-time. 

{\slshape Eq. (\ref{eq:vxc})  is the main result 
of this work}. It is a variational equation that establishes 
a causal connection between $u\XC$ in real-time with an XC action functional in the Keldysh space, 
and the memory of the system. 
If an approximation 
to the XC action functional is known, then Eq. (\ref{eq:vxc}) can be used to 
estimate the XC potential.
The potentials $u(\mb{r}t)$ and $u_{\mr{s}}(\mb{r}t)$ 
also satisfy the same type of equation as that of $u\XC$;
one has to replace $f\XC$ and $\mc{A}\XC$ by $\chi^{-1}$ and $\mc{B}$, or
$\chi_{\mr{s}}^{-1}$ and $\mc{B}_{\mr{s}}$.

Note that the left-hand side of Eq. (\ref{eq:vxc}) is a functional of the density $\bar{n}\cc(\mb{r}t^{\pm})$, 
or simply $n(\mb{r}t)$. This implies that the second functional derivative of $\mc{A}\XC$ with respect to the 
density in real-time is not 
symmetric, i.e.:
\begin{equation}
\frac{\delta }{\delta n(\mb{r}'t')}\frac{\dbar \mc{A}\XC}{\dbar n(\mb{r}t)}=0\quad t'\ge t~.
\end{equation}
Here, the symbol $\delta /\delta n(\mb{r}'t')$ represents regular functional differentiation 
in real-time because the operation $\dbar/\dbar n(\mb{r}t)$ already involves evaluation at the 
physical regime. 
The above result is a consequence of implementing causality in the $\mc{C}$ space explicitly using 
the path-ordering operator. 
Furthermore, recursive differentiation of Eq. (\ref{eq:vxc}) also allows us to express 
its solution as a series of functional derivatives of $\mc{A}\XC$. This reads
\begin{equation}
u\XC(\mb{x}_1)=\frac{\dbar \mc{A}\XC}{\dbar n(\mb{x}_1)}+w\XC(\mb{x}_1)
\een
where
\ben
w\XC(\mb{x}_1)=\sum_{m=2}^{\infty}
\frac{(-1)^{m+1}}{m!}\int \ud\mu(\mb{x}_2)\cdots \ud\mu(\mb{x}_m) \frac{\delta^{m-1}}
{\delta n(\mb{x}_m)\cdots \delta n(\mb{x}_2)} \frac{\dbar \mc{A}\XC}{\dbar n(\mb{x}_1)}~.
\end{equation}
Here $\mb{x}_m=\mb{r}_m,t_m$, $m=1,2,\ldots$ and $\ud \mu(\mb{x}_m)=n(\mb{x}_m)\ud^4\mb{x}_m$. 
The functional derivatives in the integral are zero if, for any $i>j$, $t_i\ge t_j$.
This series shows that the XC potential depends on perturbations
of the XC potentials in all orders. However, in order to achieve convergence 
the functional derivatives must decrease as their order increases.

Now let us apply our variational equation to the derivation of the ALDA XC potential:
\begin{equation}
\mc{A}\XC^{\mr{ALDA}}[n\cc]=\int_{\zi}^{\zf}\ud z \int \ud^3\mb{r}
~[\epsilon\XC(n)n]\Bigg|_{n=n\cc(\mb{r}z)}~,
\end{equation}
where $\epsilon\XC$ is the local XC energy density. To solve Eq. (\ref{eq:vxc})
the memory term can be neglected to yield
\begin{equation}
\bar{u}\cXC^{\mr{ALDA}}(\mb{r}z)=\frac{d }{d n}[\epsilon\XC(n)n]\Bigg|_{n=n\cc(\mb{r}z)}~.
\end{equation}
Further differentiation leads to the kernel formula:
\begin{equation}
f\XC^{\mr{ALDA}}(\mb{r}t,\mb{r}'t')=\delta(\mb{r}-\mb{r}')\delta (t-t')\frac{d^2 }{d n^2}
[\epsilon\XC(n)n]\Bigg|_{n=n(\mb{r}t)}~.
\end{equation}
The singularity of the XC kernel does not contribute to the integral term of Eq. (\ref{eq:vxc}) 
because the end point is not included, or in other words, the end point is approached in a limiting procedure.
Hence, the above equation satisfies Eq. (\ref{eq:vxc}) and thus it is the solution of it.
The singularity of the XC kernel arises from the definition of the XC potential, which implies that
at equal-times the XC kernel must cancel the singularity of the Hartree kernel. However, the ALDA XC
kernel does not cancel the singularity of the Hartree kernel due to the self-interaction error. 

Another application is the TDOPM. The exchange functional form
remains the same as the one proposed by \citet{L98}:
\begin{equation}
\mc{A}_{\mr{x}}[n\cc]=\int_{\zi}^{\zf} \ud z~\langle \Phi\cc[n\cc](z)|\hat{W}|\Phi\cc[n\cc](z)\rangle
-\mc{A}\Ha[n\cc]~~.
\end{equation}
To derive the TDOPM one has to assume that (for example, see \cite{U12}):
\begin{equation}
\frac{\delta \mc{A}_{\bar{v}\cc}}{\delta n(\mb{r}z)}=
\frac{\delta \mc{A}_{\mr{s},\bar{v}\cs}}{\delta n(\mb{r}z)}~.
\end{equation}
If we set $\mc{A}_{\mr{xc}}=\mc{A}_{\mr{x}}$
and expand $A_{\bar{v}\cc}$ using Eq. (\ref{expansion}) we find that
the memory term in Eq. (\ref{eq:vxc}) can be discarded. Hence we can write:
\begin{equation}
u_{\mr{x}}(\mb{r}t)=\frac{\dbar \mc{A}_{\mr{x}}}{\dbar n(\mb{r}t)}~.
\end{equation}
The right hand side of the above equation can be calculated using 
the chain rule. If the result is multiplied by $\chi_s$ and then integrated,
the final result coincides with that of \citet{UGG95}.

Ground-state DFT is also accessible with this theory. 
We can introduce a slowly varying density $n\cc^T(\mb{r}z)=n\cc(\mb{r}z/T)$, where $T\rightarrow\infty$. 
One can use the adiabatic theorem to show that:
\begin{equation}
\lim_{T\rightarrow \infty}A\XC[n\cc^T]=\lim_{T\rightarrow\infty} \int_{\zi}^{\zf}
\ud z~ E\XC[n\cc^T(\cdot,z)]~~,
\end{equation}
where $E\XC$ is the XC energy functional of DFT.
The above equation is local in time. As in the previous case, 
the solution of Eq. (\ref{eq:vxc}) has to be of the form:
\begin{equation}
\lim_{T\rightarrow \infty} u\XC[n^T](\mb{r}t)=
\lim_{T\rightarrow \infty}\frac{\delta E\XC}{\delta n(\mb{r})}\Bigg|_{n=n^T(\mb{r}t)}~~.
\end{equation}
where $n^T=\bar{n}\cc^T$.

\section{Conclusions}

To summarize, we examined the RG action functional and the solution of the causality paradox by \citet{V08}
from the point of view of unitary evolution. We extended this solution to the Keldysh space, and, 
under the strict causality assumption, we found a variational equation for the XC 
potential that involves an XC memory term. 
The solution of this variational equation is a series in terms of 	functional 
derivatives of the XC action functional in the Keldysh space. 
We showed that it is possible to derive the 
ALDA XC and TDOPM exchange potentials from the present theory and that ground states are 
also accessible using the adiabatic theorem. 

\section{Acknowledgements}
The author thankfully acknowledges valuable discussion with Adam Wasserman, Daniel Jensen, and Daniel Whitenack.

\bibliography{vtddft}

\end{document}